\documentclass[preprint,12pt]{elsarticle}

\usepackage{amssymb}
\usepackage{amsmath}
\usepackage{dsfont}
\usepackage{bm}
\usepackage{xcolor}
\usepackage{algpseudocode} 
\usepackage{algorithm}
\usepackage{tabularx}
\usepackage{placeins}
\usepackage{subfig}
\usepackage{adjustbox}

\newtheorem{definition}{Definition}
\newtheorem{assumption}{Assumption}

\journal{Journal of Power Sources}

\begin{document}

\begin{frontmatter}




\title{Battery Capacity Knee-Onset Identification and Early Prediction Using Degradation Curvature}


\author[label1,label2]{Huang Zhang\corref{cor1}}
\ead{huangz@chalmers.se}
\author[label2]{Faisal Altaf}
\ead{faisal.altaf@volvo.com}
\author[label1]{Torsten Wik}
\ead{torsten.wik@chalmers.se}
\cortext[cor1]{Corresponding author.}
\affiliation[label1]{organization={Department of Electrical Engineering},
            addressline={Chalmers University of Technology}, 
            city={Gothenburg},
            postcode={41296}, 
            country={Sweden}}
            
\affiliation[label2]{organization={Department of Electromobility},
            addressline={Volvo Group Trucks Technology}, 
            city={Gothenburg},
            postcode={40508}, 
            country={Sweden}}

\begin{abstract}
Abrupt capacity fade can have a significant impact on performance and safety in battery applications. 
To address concerns arising from possible knee occurrence, this work aims for a better understanding of their cause by introducing a new definition of capacity knees and their onset.
A curvature-based identification of a knee and its onset is proposed, which relies on the discovery of a distinctly fluctuating behavior in the transition between an initial and a final stable acceleration of the degradation. The method is validated on experimental degradation data of two different battery chemistries, synthetic degradation data, and is also benchmarked to the state-of-the-art knee identification method in the literature.
The results demonstrate that our proposed method could successfully identify capacity knees when the state-of-the-art knee identification method failed. Furthermore, a significantly strong correlation is found between knee and end of life (EoL) and almost equally strong between knee onset and EoL.
As the method does not require the full capacity fade curve, this opens up online knee-onset identification as well as knee and EoL prediction.

\end{abstract}


\begin{highlights}
\item A curvature approximation method to measure the rate of capacity fade is proposed. 
\item A new oscillatory degradation phenomenon is discovered.
\item The oscillatory degradation phenomenon is used to identify knee and knee-onset.
\item The method outperforms the state-of-the-art method on both synthetic and real data.
\item The method can be used for online knee-onset identification and knee prediction.
\end{highlights}

\begin{keyword}
Battery diagnosis \sep Degradation mechanisms \sep Knee identification \sep Knee-onset early prediction \sep Curvature approximation


\end{keyword}

\end{frontmatter}


\section{Introduction}
Lithium-ion batteries have been widely used as energy storage systems in various applications, such as electric vehicles and microgrids, due to their high power and energy density, rapid response, and long lifetime characteristics \cite{schmuch2018performance}. However, as a result of a complex interplay of different physical and chemical degradation mechanisms, the performance (e.g., available energy and available power) of lithium-ion batteries gradually degrades over their service lives, where the degradation rate is a nonlinear function of storage and cycling conditions (temperature, state-of-charge (SoC) window, charge/discharge current, energy throughput, etc). In some cases, accelerated capacity fade can not only lead to accelerated performance degradation but also battery safety issues \cite{martinez2018technical}.

In both field data and experimental testing of commercial lithium-ion batteries, it has been observed that the capacity fade often exhibits a two-stage behavior, with a slow degradation rate in the first stage and then an accelerated degradation rate in the second stage \cite{dubarry2012synthesize} \cite{he2011prognostics} \cite{yang2017prognostics}. The transition from the first stage to the second stage infers a knee appearance on the capacity fade curve. Generally, the knee occurs within a 70-95\% capacity retention window depending on the operating conditions \cite{attia2022knees}. Furthermore, it has been experimentally demonstrated that reusing lithium-ion batteries in less-demanding second-life applications may not slow down the aging trend once the knee has already occurred \cite{martinez2018technical} \cite{martinez2016evaluation} \cite{zhang2019accelerated}. Therefore, for safety and performance reasons, the occurrence of the knee should be avoided, or at least delayed, to ensure a long battery lifetime. Batteries with knee occurrence should normally be retired immediately from operation, no matter if it is in their first-life or later-life application \cite{martinez2018technical}.

It is common for manufacturers of electric vehicles to provide a battery warranty of 8-10 years, guaranteeing 70-80\% of their initial nominal capacity \cite{wood2011investigation} \cite{arrinda2021application}. When a battery pack that consists of thousands of battery cells retires from its first life in an electric vehicle, not all battery cells in the battery pack have necessarily reached 70-80\% of their initial nominal capacity, and knee at cell level may occur before or after the end of life defined at pack level, due to cell-to-cell variations in the pack \cite{baumann2018parameter}. Instead of being recycled or disposed of, one desirable option is to repurpose retired batteries to less demanding second-life applications as stationary battery energy storage systems (BESSs) \cite{ahmadi2017cascaded}. However, to ensure the lifetime is maximized in its second-life application, a knee occurrence on the capacity fade curve needs to be identified prior to repurposing the retired batteries for a second-life application.

To date, several studies have attempted to firstly define the knee and then propose a method to identify it, both in off-line scenarios \cite{diao2019algorithm} \cite{fermin2020identification} \cite{greenbank2021automated}, and online scenarios \cite{zhang2019accelerated} \cite{sohn2022two} \cite{costa2024icformer}. 
Diao et al. \cite{diao2019algorithm} defined the knee as the intersection of two tangent lines at two points (i.e., the points with minimum and maximum absolute slope, respectively) on the capacity fade curve, and then developed an empirical degradation model to characterize the capacity fade curve from experimental data, on which the two points were identified to locate the two tangent lines. However, different degradation models may be required to fit different types of capacity fade curves. 
Ferm{\'\i}n-Cueto et al. \cite{fermin2020identification} also defined the knee as the intersection of two straight lines identified by directly fitting the Bacon Watts model to capacity degradation data. The Bacon-Watts method is simple and robust against noise without superimposing a degradation model, but may not be applicable to all types of capacity fade curves.
The bisector method, proposed by Greenbank and Howey \cite{greenbank2021automated}, first fits the early and late life capacity fade gradients using linear regression. Then the knee is identified as the intersection of an angle bisector of two gradients and the capacity fade curve. However, the bisector method is sensitive to the selection of early and late life capacity fade data, and may therefore not be applicable to all types of capacity fade curves.
Zhang et al. \cite{zhang2019accelerated} firstly learned a strip-shaped safety zone from experimental data of the height of a peak on the incremental capacity curve, and then the knee was identified as the last cycle of four consecutive cycles beyond the safety zone using quantile regression and Monte Carlo simulation. Although the quantile regression method works with incoming data streams, the identified knees vary with the amount of available online data.  
Sohn et al. \cite{sohn2022two} proposed a convolutional neural networks model to extract temporal features from time-series data to predict the number of cycles left to the knee point. However, their method requires extensive knee labeling beforehand for the model training process using the Bacon-Watts model.
In a very recent work by Costa et al. \cite{costa2024icformer}, a transformer-based deep learning model integrated with incremental capacity analysis was proposed for knee identification and degradation diagnosis. The knee identification was formulated into a binary classification problem, i.e., the model predicts whether or not a knee will occur within a window size of 800 cycles. Although the model can quantify degradation modes in addition to the knee indication within 800 cycles, it also requires knee labeling beforehand and does not identify the exact knee point on the capacity fade curve, which is often preferable in practical applications, such as in the classification of retired electric vehicle batteries.
To summarize, the aforementioned knee identification methods can be divided into two categories. One is based on finding the intersection between a straight line approximating the early fade and a line following the rate of fade after a knee \cite{diao2019algorithm} \cite{fermin2020identification} \cite{greenbank2021automated}. The disadvantages of this approach are that, depending on the shape of the capacity fade curve, they may fail, and they cannot be used for online identification and prediction as they need more or less the complete fade curve. The other category is methods based on machine learning models \cite{zhang2019accelerated} \cite{sohn2022two} \cite{costa2024icformer}. These methods can be used for online identification and prediction, but they require large amounts of labeled data and other inputs than only capacity.

\textbf{Contribution:}
The objective of this work is to fill in the gap indicated above by proposing a generalized capacity knee identification method that leverages battery degradation prior knowledge to improve knee identification performance. We formulate the capacity knee identification problem as an unsupervised time series segmentation problem given an assumption of three consecutive discrete states of the degradation process, from the beginning of life till the end of life. Our key results and contributions are:     
\begin{enumerate}
\item
We are the first to propose the use of approximated curvature to measure the rate of change of capacity fade rate in discrete time, and this reveals a new oscillatory degradation phenomenon. This degradation phenomenon occurs during a period that clearly separates the capacity fade curve into three distinct intervals that can be automatically identified offline as well as online with some delay. The knee onset corresponds to the start of the oscillatory middle interval and the actual knee corresponds to the end of this period.
\item We validate the effectiveness of our proposed method on two experimental datasets of two different cathode chemistries (i.e., LFP and NMC) and one synthetic dataset, the three datasets covering knee occurrence induced by lithium plating, resistance growth, and particle cracking. Compared to the state-of-the-art method (i.e., the double Bacon-Watts model), our proposed method consistently identifies knee and knee-onset points on capacity fade curves over all three battery datasets while the double Bacon-Watts model completely fails on one of the battery datasets.
\item
With strong correlations between knee and knee-onset (identified using our proposed identification method) and the end of life, the knee-onset can provide an early warning of accelerated degradation. Therefore, we demonstrate that learning early-prediction models for battery knee-onset prediction can significantly reduce experimental time and its associated costs. Moreover, knee-onset prediction can potentially benefit several industrial applications with significant economic value (e.g., validation of formation protocols, battery grading, battery replacement planning, and battery repurposing for second-life applications).
\end{enumerate}

\section{Background} \label{sec2}
\subsection{Double Bacon-Watts model}
In order to identify both knee and knee-onset points on capacity fade curves instead of only one knee point, the double Bacon-Watts model is adapted from the Bacon-Watts model \cite{fermin2020identification}, i.e.,
\begin{equation}\label{eq:0}
\begin{aligned}
    Y = & \alpha_0 + \alpha_1(x-x_0) + \alpha_2(x-x_0)\mathrm{tanh}\{ (x-x_0)/\gamma \} \\
        & + \alpha_3 (x-x_2)\mathrm{tanh}\{ (x-x_2)/\gamma \} + Z,
\end{aligned}
\end{equation}
where $\alpha_0$ is an intercept; $\alpha_1$, $\alpha_2$, and $\alpha_3$ denote the slopes of the intersecting lines; $\gamma$ determines the abruptness of the transition and is set to a low value (i.e., $10^{-8}$) to achieve an abrupt transition; $Z$ is a zero mean, normally distributed random variable; $x_0$ and $x_2$ are two transition points, i.e., knee-onset and knee.

Given a sequence of measured capacity data of a cell, the Levenberg–Marquardt nonlinear least squares algorithm is used to learn the model parameters in Eqn. (\ref{eq:0}). Moreover, initial values for the model parameters that are used in this work are given in Table \ref{tab1}. Note that $N$ is the number of sampled capacity points.
\begin{table}[!htbp]
\renewcommand{\arraystretch}{1.5}
\caption{Initial values for the double Bacon-Watts model parameters}
\begin{center}
\begin{tabular}{|c|l|c|c|}
\hline
\textbf{Parameter} & \textbf{Description} & \textbf{Value} & \textbf{Unit}\\
\hline
$\alpha_0$ & Intercept & 1 & Ah\\
\hline
$\alpha_1$ & Slope of the intersecting line 1 & $-10^{-4}$ & Ah/cycle\\
\hline
$\alpha_2$ & Slope of the intersecting line 2 & $-10^{-4}$ & Ah/cycle\\
\hline
$x_0$ & knee-onset point & $\frac{7N}{10}$ & cycle \\
\hline
\end{tabular}
\label{tab1}
\end{center}
\end{table}

\subsection{Matrix profile}\label{subsec2_1}
Here, we first introduce all the necessary definitions related to matrix profile \cite{yeh2016matrix}, which will be used later in our proposed method:
\begin{definition}
A time series $\bm{Y} = \{y_{d_1}, y_{d_2}, ..., y_{d_M}\}$ is an ordered sequence of real values that are evenly spaced, where $M$ is the length of $\bm{Y}$.
\end{definition}
\begin{definition}
A subsequence $\bm{Y}_{j,L} \subseteq \bm{Y}$ is a continuous subset of the values from $\bm{Y}$ of length $L$ starting from $j$, i.e., $\bm{Y}_{j,L} = \{y_{d_j}, y_{d_{j+1}}, ..., y_{d_{j+L-1}}\}$, where $1 \leq j \leq M-L+1$.
\end{definition}
\begin{definition}
The set $\bm{Z}$ of a time series $\bm{Y}$ is an ordered set of all subsequences of $\bm{Y}$ obtained by sliding a window of length $L$ across $\bm{Y}$ such that: $\bm{Z}=\{\bm{Y}_{1, L}, \bm{Y}_{2, L}, ..., \bm{Y}_{M-L+1, L}\}$, where $L$ is a user-defined subsequence length.
\end{definition}
\begin{definition}
A self-similarity set $\bm{J}_{\bm{ZZ}}$ is a set containing pairs of each subsequence $\bm{Y}_{j,L}$ in $\bm{Z}$ with its nearest neighbor $\bm{Y}_{k,L}$ in $\bm{Z}$. We denote this as $\bm{J}_{\bm{ZZ}} = \{ (\bm{Y}_{j,L}, \bm{Y}_{k,L}) \}, j \neq k$.
\end{definition}
\begin{definition}
A matrix profile $\bm{P}_{\bm{ZZ}}$ is a vector of the Euclidean distances between the two subsequences of each pair in $\bm{J}_{\bm{ZZ}}$.
\end{definition}
\begin{definition}
A matrix profile index $\bm{I}_{\bm{ZZ}}$ of a self-similarity set $\bm{J}_{\bm{ZZ}}$ is a vector of integers where $\bm{I}_{\bm{ZZ}}[j] = k$ if $(\bm{Y}_{j,L}, \bm{Y}_{k,L}) \in \bm{J}_{\bm{ZZ}}$.
\end{definition}

\subsection{Scalable time series anytime matrix profile}
To efficiently compute these two time series, i.e., the matrix profile $\bm{P}_{\bm{ZZ}}$ and matrix profile index $\bm{I}_{\bm{ZZ}}$, the algorithm that we adopt in this work is called Scalable Time series Anytime Matrix Profile (STAMP) \cite{yeh2016matrix}. STAMP is a time-series all-pairs one-nearest-neighbor search algorithm that uses the Fast Fourier Transform for speed and scalability. There are only two input parameters, i.e., the time series $\bm{Y}$, and a subsequence length $L_1$, where $L_1$ is the desired length of the time series pattern to search for. 

\subsection{Arc curve}
Here we introduce all the necessary definitions related to the arc curve, which represents the likelihood of a regime change at each location \cite{gharghabi2017matrix}.
\begin{definition}
An arc is an entry pair $(j,k)$ drawn from the $j$-th entry in the matrix profile index $\bm{I}_{\bm{ZZ}}$ to its nearest neighbor location at index $k$.
\end{definition}
\begin{definition}
The Arc Curve (AC) for a time series $\bm{Y}$ of length $M$ is itself a time series of length $M$ containing nonnegative integer values. The $j$-th index in the AC specifies the number of nearest neighbor arcs from the matrix profile index that cross over the location $j$.
\end{definition}

In addition to low values at the location of the regime change, the AC also has low values at both the leftmost and rightmost edges, due to the fact that there are fewer arcs that can cross a location at the edges. In order to reduce the edge effect on the AC, the Corrected Arc Curve (CAC) is often calculated instead.

\subsection{Fast low-cost unipotent semantic segmentation}
To produce the third time series, i.e., the Corrected Arc Curve (CAC), we adopt the Fast Low-cost Unipotent Semantic Segmentation (FLUSS) algorithm \cite{gharghabi2017matrix}. FLUSS (see Algorithm \ref{algorithm_1}) has only two input parameters, i.e., the matrix profile index $\bm{I}_{\bm{ZZ}}$ and the subsequence length $L_2$.

\begin{algorithm}
	\caption{FLUSS} \label{algorithm_1}
	\begin{algorithmic}[1]
	    \State \textbf{Input:} 1. $\bm{I}_{\bm{ZZ}}$: matrix profile index with the length of $M-L_2+1$.
	    \State \textbf{Input:} 2. $L_2$: length of the subsequence
	    \State Initialize three empty arrays with lengths of $M-L_2+1$, i.e., $\mathrm{AC}$, $\mathrm{CAC}$, $\mathrm{MARK}$.
		\For {i=1:$M-L_2+1$}
            \State Count the number of arcs that cross over index $i$ and then store in $\mathrm{MARK}$
		\EndFor
        \For {i=1:$M-L_2+1$}
            \State Cumulatively sum values in $\mathrm{MARK}$ for each index $i$ and store in $\mathrm{AC}$
        \EndFor
        \State $\mathrm{IAC}= \text{parabolic curve of length } n \text{ and height } n/2$ \cite{gharghabi2017matrix}
        \State $\mathrm{CAC}=\mathrm{min}(\mathrm{AC}/\mathrm{IAC}, 1)$
	    \State \textbf{Output:} $\mathrm{CAC}$
	\end{algorithmic} 
\end{algorithm}

\subsection{Regime extracting algorithm}
With the advantage of having only two input parameters, i.e., the subsequence length $L_2$, and the number of states $N_{S}$, we adopt the regime extracting algorithm (REA) \cite{gharghabi2017matrix} to extract the locations of the state changes from the CAC, i.e., $\bm{B}$. 

\begin{definition}
A battery cell health degradation process with knee occurrence on the capacity fade curve consists of three discrete states $\bm{S} = [s_1, s_2, s_3]$ separated by two boundaries $\bm{B} = [b_1, b_2]$. Here, $s_1$ represents the cell degradation process from the beginning of life to the knee-onset point, $s_2$ represents the cell degradation process from the knee-onset point to the knee point, and $s_3$ represents the cell degradation process from the knee point to the end of life.
\end{definition}

\begin{algorithm}
	\caption{Regime Extracting Algorithm} \label{algorithm_2}
	\begin{algorithmic}[1]
	    \State \textbf{Input:} 1. CAC: corrected arc curve for approximated
	    curvature
	    \State \textbf{Input:} 2. $L_2$: length of the subsequence
	    \State \textbf{Input:} 3. $N_{S}$: number of state changes 
	    \State Initialize an empty array $\bm{B}$ of length $N_{S}$
		\For {i=1:$N_{S}$}
            \State $\bm{B}(i) = \mathrm{index}(\mathrm{min}(\mathrm{CAC}))$
            \State Set an exclusion zone as five times the subsequence length $L_2$ before and after $\bm{B}(i)$
		\EndFor
	    \State \textbf{Output:} $\bm{B}$
	\end{algorithmic} 
\end{algorithm}

\section{Methods}
Researchers from different areas come across knee identification problems, in which knees can be detected in either an ad-hoc manner or with a general tool \cite{satopaa2011finding}. The concept of a knee in the battery field generally relates to the degradation rate on the capacity fade curve. The degradation rate in terms of capacity fade is the result of the convolution of various underlying degradation mechanisms and possible interactions between them. Extrinsic factors, such as the sequence of aging tests, may influence the degradation rate in terms of capacity fade, especially at high C-rates and long duration of continuous cycling \cite{raj2020investigation}. Intrinsic factors, such as battery chemistry, and manufacturing variances, also have an impact on the degradation rate \cite{bach2016nonlinear}. To design a general knee identification method, a consistent knee definition that is applicable to batteries of any chemistry and a wide range of operating conditions is required. 

To measure the rate of change of the capacity fade, we first introduce the concept of curvature, which is a mathematical measure of the amount by which a curve deviates from being a straight line \cite{satopaa2011finding}. For a continuous function $f$, the curvature $\kappa(x)$ of $f$ at any point, is defined as
\begin{equation}\label{eq1}
    \kappa(x) = \frac{f''(x)}{(1+f'(x)^2)^{1.5}}.
\end{equation}
The curvature value calculated at one point using Eqn (\ref{eq1}) can be positive, negative, or 0, depending on the second derivative of the function $f$. 

Although a knee can be mathematically well-defined as the point of maximum curvature for continuous functions \cite{satopaa2011finding}, it is in practice challenging to accurately identify the knee using Eqn (\ref{eq1}) as the capacity fade data is sampled and noisy. Therefore, the curvature needs to be approximated before the knee identification on discrete data.
\subsection{Degradation curvature approximation}
\begin{figure*}[!ht]
\centering
\subfloat[The capacity fade data points.]{\includegraphics[width=2.5in]{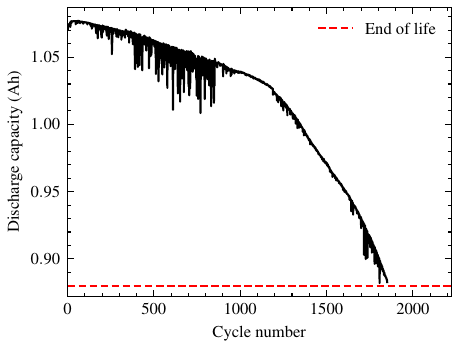}%
\label{fig1_1}}
\hfil
\subfloat[The time series approximated curvature.]{\includegraphics[width=2.5in]{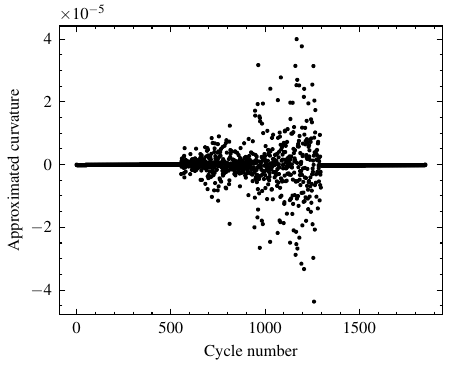}%
\label{fig1_2}}
\caption{The curvature approximation results of a sample cell [b1c0] in the Toyota Research Institute dataset \cite{severson2019data}.}
\label{fig1}
\end{figure*}

The starting point is a sequence of measured capacity data from a lithium-ion cell, $\{x_i, y_i\}_{i=1}^N$, where $x_i \in \mathbb{R}_0^+$ is the number of cycles that the battery has been used, $y_i \in \mathbb{R}^+$ is the discharge capacity measured per cycle, and $N$ is the number of sampled capacity points in the set. As an example, a set of discrete capacity fade data points of a sample cell, [b1c0] from the Toyota Research Institute dataset \cite{severson2019data}, is illustrated in Fig. \ref{fig1_1}. 




Our curvature-based method relies on the following two assumptions:
\begin{assumption}
The lithium-ion cell has a knee occurrence on its capacity fade curve before the end of the experiment.
\end{assumption}
\begin{assumption}
The $x$-values are evenly spaced. If not, the data points are fitted to a spline function and interpolated to become so.
\end{assumption}

We summarize the proposed curvature approximation in a step-by-step manner as follows: 
\begin{enumerate}
    \item To have our capacity knee identification method as little affected as possible by variations in battery capacity magnitude, the raw capacity data points $\{x_{i}, y_{i}\}_{i=1}^N$ are first normalized with the battery's initial nominal capacity $Q_{nom}$. The resulting set of normalized data points is $\{x_i , y_{{n}_i}\}_{i=1}^N$ where
    \begin{align*}
        y_{{n}_i} &= \frac{y_{i}}{Q_{nom}}.
    \end{align*}
    \item As a low-pass filter that utilizes the local least-square polynomial approximation, the Savitzky-Golay filter provides competitive denoising performance \cite{savitzky1964smoothing}. Since it is more computationally efficient than many other smoothing techniques with potential for real-time applications, it is used to smooth the normalized data points $\{x_i, y_{{n}_i}\}_{i=1}^N$. The resulting set of smoothed data points $\{x_i, y_{{sn}_i}\}_{i=1}^N$ is then used in the next step.
    \item We approximate the curvature at data points $\{x_i, y_{{sn}_i}\}_{i=(ws+1)/2)}^{N-(ws-1)/2}$ by calculating their corresponding successive differences $y_{d_i}$ 
    where 
    \begin{align*}
        y_{d_i} &= y_{{sn}_{i-(ws-1)/2}} + y_{{sn}_{i+(ws-1)/2}} - 2y_{{sn}_{i}}.
    \end{align*}
    With the assumption that the $x$-values are evenly spaced, then $y_{d_i} = 0$ for any straight line. However, if any consecutive three points form a knee, then $y_{d_i} < 0$ as the middle point $y_{{sn}_{i}}$ is now above the straight line that goes through the first point $y_{{sn}_{i-(ws-1)/2}}$ and the third point $y_{{sn}_{i+(ws-1)/2}}$. Analogously, if any consecutive three points form an elbow, then $y_{d_i} > 0$.
\end{enumerate}

The resulting approximated curvature of the example cell in Fig. \ref{fig1_1} is illustrated in Fig. \ref{fig1_2}.
Contrary to what one would perhaps expect from Fig. \ref{fig1_1}, having most fluctuations at the beginning, we observe a very stable start of the curvature followed by a distinct period of large fluctuations, eventually followed by an abrupt change to yet another stable period.

As a matter of fact, drastic changes of system states from one stable state to another stable state through a critical transition state can also be observed in other complex dynamical systems in ecology, biology, economics, and other fields \cite{scheffer2009early} \cite{liu2015identifying}. Since we know that the fade curve exhibits accelerated aging in the last phase and a stable degradation (after the low-pass filtering) in the first phase, it seems rational to regard the initiation of the transition phase as the knee onset and then define the end of this phase as the knee point.

Next, we will use the algorithms in Section \ref{sec2} to automate the process of identifying the state of transition phase.

\subsection{Knee and knee-onset identification}
As suggested by~\cite{yeh2016matrix}, it is empirically shown that given the matrix profile and the matrix profile index, the resulting corrected arc curve contains information about a possible regime change at each location of a time series. Therefore, we employ a similar approach to knee and keen-onset identification. Specifically, our goal is to obtain three time series, i.e., matrix profile, matrix profile index, and corrected arc curve, to represent a time-series approximated curvature that is produced in the previous section.

\begin{figure*}[!ht]
\centering
\subfloat[The time series approximated curvature (top) and corresponding arc curve (bottom).]{\includegraphics[width=2.5in]{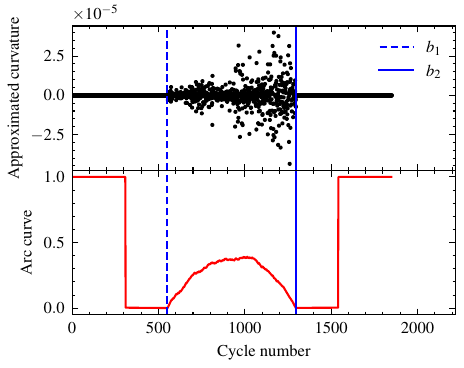}%
\label{fig2_1}}
\hfil
\subfloat[Identified capacity knee and capacity knee-onset.]{\includegraphics[width=2.5in]{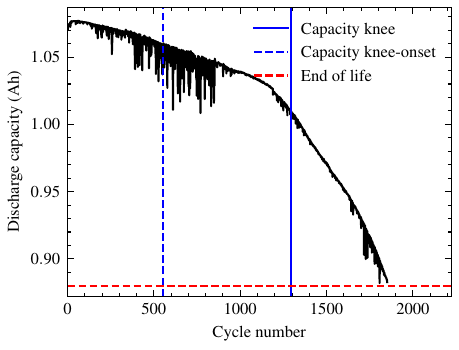}%
\label{fig2_2}}
\caption{The knee-onset and knee identification results of the sample cell [b1c0].}
\label{fig2}
\end{figure*}

In Fig. \ref{fig2_1}, the time series approximated curvature of the sample cell [b1c0] is shown at the top, while its corresponding CAC is shown at the bottom. All the parameters that are used to produce Fig. \ref{fig2_1} are listed in Table \ref{tab2}. 
It can be seen from Fig. \ref{fig2_1} that the approximated curvature is approximately equal to zero from the beginning of life till the first state change point $b_1$, which is identified as the capacity knee-onset using REA. Until the second state change point $b_2$, the approximated curvature fluctuates significantly, which indicates that the rate of change of degradation rate on the capacity fade curve fluctuates. After the second state change point $b_2$ that is identified as the capacity knee using REA, the degradation rate accelerates on the capacity fade curve till the end of life, as shown in Fig. \ref{fig2_2}.

Intuitively, the time series segmentation algorithm using its corresponding CAC in the context of a battery capacity fade process is straightforward. For instance, suppose the time series approximated curvature $\bm{Y}$ of the sample cell [b1c0] has a state change at location $a$, we would expect very few arcs to cross over $a$ as most of the subsequences $\bm{Y}_{j, L}$ will find their nearest neighbors within the same state. Therefore, the height of the CAC should be the lowest at the location of the boundary where the state changes from one to another. 

\begin{table}[!htbp]
\renewcommand{\arraystretch}{1.5}
\caption{Input parameters in the proposed method}
\begin{center}
\begin{tabular}{|c|l|c|c|}
\hline
\textbf{Parameter} & \textbf{Description} & \textbf{Value} & \textbf{Unit}\\
\hline
$ws$ & \shortstack[l]{Sliding window size to approximate \\curvature} & 3 & cycle\\
\hline
$L_1$ & \shortstack[l]{Subsequence length to produce matrix \\profile} & 3 & cycle\\
\hline
$L_2$ & Subsequence length to produce CAC & $[\frac{N}{5}]$ & cycle\\
\hline
$N_S$ & \shortstack[l]{Number of states to identify knee and \\ knee-onset} & 3 & - \\
\hline
\end{tabular}
\label{tab2}
\end{center}
\end{table}

\subsection{Experimental design}
\subsubsection{Toyota research institute dataset}
\paragraph{Dataset description}
The first battery dataset used in this work was generated by Toyota Research Institute in collaboration with Stanford University and Massachusetts Institute of Technology \cite{severson2019data} \cite{attia2020closed}, in which there are 169 lithium iron ferrous phosphate (LFP)/graphite cylindrical cells manufactured by A124 Systems (model APR18650M1A, 1.1 Ah nominal capacity). The total 169 cells are from 4 batches ("2017-05-12", "2017-06-30", "2018-04-12", and "2019-01-24") with batch name denoting the date when the experiment was started. 
The cells were charged with a one-step or multi-step fast-charging protocol from 0\% to 80\% SoC and then charged with a uniform 1C constant current–constant voltage (CC-CV) charging step from 80\% to 100\% SoC. Subsequently, cells were discharged identically at 4C rate to 0\% SoC. All cells were tested in an environmental chamber at a constant ambient temperature of 30$^{\circ}$C. 
The cells were cycled until they reached the end of life (EoL) threshold, set to 80\% of their initial nominal capacity. 
Time-series cell voltage, current, and (surface) temperature in each cycle were continuously measured, while two battery health metrics, i.e., rated capacity (4C discharge, 30$^{\circ}$C) and internal resistance ($\pm3.6$ C pulse current, 30 or 33 ms pulse width, 80\% SoC) were measured per cycle. 
All the cells in this dataset have knee occurrence on their capacity fade curves before the end of the experiment. 

\paragraph{Lithium plating-induced knees}
The knee occurrence in this dataset is caused by lithium plating due to loss of delithiated negative electrode active material. Specifically, at high rates of loss of delithiated negative electrode active material (LAM), the negative electrode capacity eventually falls below the remaining lithium inventory in a cell. Then, the negative electrode will not be able to accommodate all the lithium from the positive electrode during charge, which leads to irreversible lithium plating and a resulting loss of lithium inventory (LLI). The loss of delithiated negative electrode active material, together with LLI, contribute to accelerated capacity fade, at which a knee also occurs. Moreover, a higher charge C-rate also accelerates the occurrence of the knee in this dataset.
Lithium plating-induced knees are commonly observed or hypothesized in commercial LFP/graphite cells \cite{lewerenz2017post} \cite{ansean2017operando} \cite{lewerenz2017differential} \cite{severson2019data} \cite{attia2020closed}. 
\subsubsection{Sandia national lab dataset}
\paragraph{Dataset description}
The second battery dataset used in this work was generated by Sandia National Laboratories \cite{preger2020degradation}, in which there are 32 lithium nickel manganese cobalt oxide (NMC)/graphite cylindrical cells from LG Chem (model 18650HG2, 3Ah). 
The NMC cells were cycled at three different ambient temperatures (15$^{\circ}$C, 25$^{\circ}$C, and 35$^{\circ}$C) with different depths of discharges (0-100\%, 20-80\%, and 40-60\%) and discharge currents (0.5C, 1C, 2C, and 3C). All the NMC cells were identically charged at 0.5C rate. To reduce the effect of manufacturing variances, at least 2 cells were tested for each combination of ambient temperature, depth of discharge, and discharge current (12 combinations). The cells were cycled beyond the end of life, defined as when they reach 80\% of their initial nominal capacity.
Time-series cell voltage, current, and (surface) temperature in each cycle were continuously measured, while one battery health metric, i.e., rated capacity (0.5C discharge at the same ambient temperature as that in each cycling test) was measured periodically.
Note that 4 cells that were cycled with 40-60\% depth-of-discharge are excluded from this work due to the fact that neither the knee nor EoL (i.e., 80\% of their initial nominal capacity) was experienced before the end of the experiment; 5 cells that were cycled with 20-80\% depth-of-discharge and one cell that was cycled with 0-100\% depth-of-discharge are also excluded from this work due to the fact that their discharge capacity data is highly corrupted. In the end, 22 cells that have knee occurrence on their capacity fade curves before the end of the experiment are included in this work.

\paragraph{Resistance-induced knees}
The knee occurrence in this dataset is attributed to resistance growth that is caused by the growth of side reaction products (e.g., solid electrolyte interphase (SEI)) on the surface of the electrode \cite{preger2020degradation}. The NMC/graphite cells in this dataset have discharge voltage-capacity curves that are relatively flat at high voltage levels and relatively steep at low voltage levels. At the beginning of life, the discharge ends within the steep region of the voltage-capacity curve. However, as the overpotential increases due to resistance growth during aging, the voltage-capacity curve is pushed downwards. As a result, a cell reaches its lower cut-off voltage more quickly, and the discharge will eventually end within the flat region of the voltage-capacity curve, making the discharge capacity highly sensitive to increasing resistance. As a result, the discharge capacity fade accelerates, assuming a linear resistance growth rate, which leads to a capacity knee.
Resistance-induced knees are commonly observed or hypothesized in cells made of oxide-based cathode materials, such as NMC, as they often operate above the stability window of the electrolyte \cite{preger2020degradation} \cite{ma2019hindering}.

\subsubsection{Synthetic dataset}
\paragraph{Particle cracking-induced knees}
The knee occurrence can also be caused by particle cracking \cite{attia2022knees}. Specifically, the intercalation and deintercalation of lithium during cycling can induce alternating mechanical stress within the electrodes, which results in particle cracking. As the cracks propagate, new surfaces are created for SEI growth, which accelerates LLI. The accelerated LLI contributes to an accelerated capacity fade and a consequent knee occurrence.
Particle cracking-induced knees have been observed or hypothesized in cells with cathodes made of NMC \cite{pfrang2018long} and NCA \cite{willenberg2020high}.
\paragraph{Battery models}
In order to verify the effectiveness of our proposed method in identifying capacity knees caused by other degradation mechanisms and their interactions, synthetic battery degradation data is generated using physics-based models. 
Specifically, underlying battery states (e.g., lithium concentrations) are simulated using a Doyle-Fuller-Newman (DFN) model \cite{doyle1995use}. To produce particle cracking-induced knees, two degradation mechanisms (i.e., SEI growth and particle cracking) at the negative electrode are coupled with the DFN model in Python Battery Mathematical Modeling (PyBaMM) library \cite{sulzer2021python}.



\paragraph{Model parameters}
The DFN model parameters (i.e., electrode parameters, electrolyte parameters) are taken from Chen et al. \cite{chen2020development} for a commercial NMC 811/graphite-$\text{SiO}_\text{x}$ cylindrical cell from LG Chem (INR21700 M50, 5 Ah). The values of electrode and electrolyte parameters are listed in the supplementary information of Ref. \cite{o2022lithium}.
The parameters of the degradation models in PyBaMM are taken from multiple sources and are also listed in the supplementary information of Ref. \cite{o2022lithium}. Here, three distinct cells with their cracking rates in Paris' law being 10, 30, and 50 times the standard particle cracking rate (i.e., $3.9 \times 10^{-20}$ \cite{purewal2014degradation}) were used to generate synthetic capacity knees for validation purposes.

\paragraph{Cycling protocol}
The cells used in this work have a nominal capacity of 5 Ah with a lower voltage cut-off of 2.5 V and an upper voltage cut-off of 4.2 V. The ambient temperature is assumed to be constant at 25$^{\circ}$C. 
The cells are charged with a 1C constant current-constant voltage (CC-CV) charging to 4.2 V and a current cut-off of C/100 (50 mA) followed by a rest for 5 minutes. The cells are subsequently discharged at 1C to 2.5 V with a current cut-off of C/100 (50 mA) and then at rest for 5 minutes. 
The simulation is set to terminate if either 1200 cycles or 80\% of the initial nominal capacity is reached. 
The discharge capacity is measured by integrating discharge current over time from 100\% SoC to the cut-off voltage.

\section{Results and discussion}
As a fundamental step prior to addressing capacity knee-related battery problems, one would like to accurately identify knee-onset and knee on the capacity fade curve and then investigate the empirical relationship between knee-onset and end of life (80\% of initial nominal capacity), and between knee and end of life. In this section, we will first validate our proposed capacity knee identification method on the two experimental degradation datasets of different battery chemistry (LFP and NMC) and then on the synthetic degradation dataset. In the validation process, we will also benchmark our capacity knee identification method to the state-of-the-art knee identification method in the literature, i.e., the double Bacon-Watts model \cite{fermin2020identification} and the numerical results of knee and knee-onset identification will also be presented. Lastly, a case study will be provided to demonstrate one promising application of knee-onset identification using our proposed method on experimental data of commercial LFP batteries.
\subsection{Validation of the proposed method}
\subsubsection{Validation on Toyota research institute dataset}
\begin{figure*}[!ht]
\centerline{\includegraphics[width=5in]{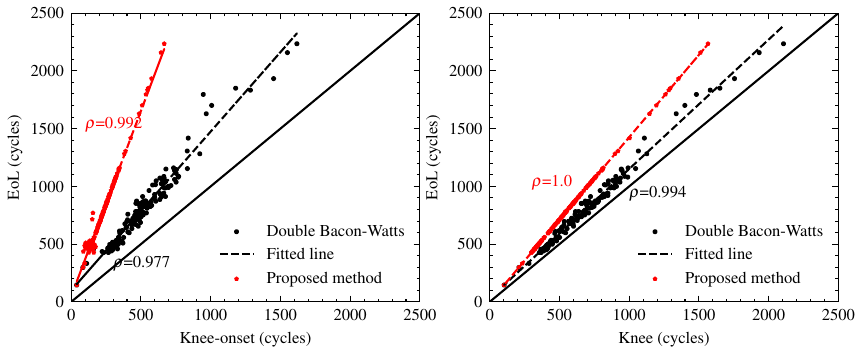}}
\caption{The relationship between knee-onset and end of life (left), knee and end of life (right) for 169 cells in the Toyota Research Institute dataset.}
\label{fig3}
\end{figure*}

With in total 169 LFP cells in the Toyota Research Institute (TRI) dataset, capacity knee-onset and capacity knee were identified for each cell using the proposed capacity knee identification method. For these, we found strong linear correlations between both knee-onset and end of life ($\rho=1.0$), and between knee and end of life ($\rho=0.992$), as shown in Fig. \ref{fig3}. 
In Table \ref{tab3}, we also compare the knee and knee-onset identification performance of our proposed method with identification using the double Bacon-Watts model. It can be concluded that the identification performance of our proposed method outperforms that of the state-of-the-art (i.e., the double Bacon-Watts model). 

Apart from the strong correlations, by referring to the diagonal line [solid black line] in Fig.~\ref{fig3}, it can be concluded that all the cells have knees occurring before the end of life. It can also be noted that the capacity knee method identified both earlier knees and earlier knee-onsets than those identified using the double Bacon-Watts model.

\subsubsection{Validation on Sandia national lab dataset}
\begin{figure*}[!ht]
\centerline{\includegraphics[width=5in]{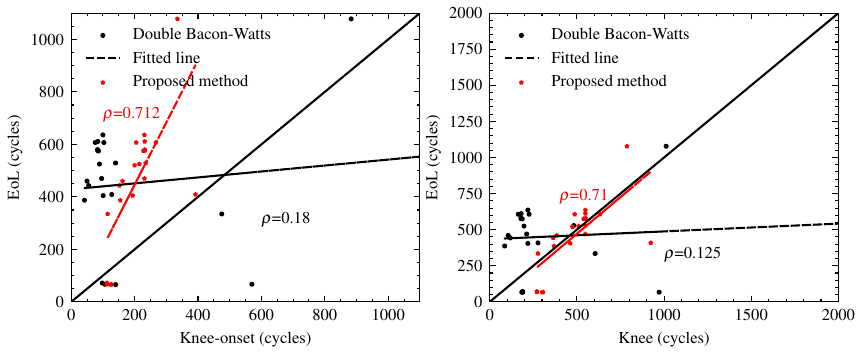}}
\caption{The relationship between knee-onset and end of life (left), and knee and end of life (right) for 22 cells in the Sandia National Lab dataset.}
\label{fig4}
\end{figure*}

With the additional 22 NMC cells in the Sandia National Lab (SNL) dataset, we again found clear linear correlations between knee-onset and end of life ($\rho=0.712$), and between knee and end of life ($\rho=0.71$) using the proposed capacity knee identification method, as shown in Fig. \ref{fig4}. However, neither the knee-onset nor the knee identified using the double Bacon-Watts model shows any clear correlations with the end of life.
As shown in Table \ref{tab3}, the knee and knee-onset identification performance of our proposed method again outperforms that of the state-of-the-art (i.e., the double Bacon-Watts model). 
\begin{table*}[!htbp]
\caption{Battery capacity knee and knee-onset identification performance}
\begin{adjustbox}{width=\textwidth,center}
\begin{tabular}{|l|c|c|c|c|}
\hline
\textbf{Method} & \multicolumn{2}{|c|}{\shortstack{\textbf{Knee identification performance} \\ \textbf{measured by Pearson's r}}} & \multicolumn{2}{|c|}{\shortstack{\textbf{Knee-onset identification performance} \\ \textbf{measured by Pearson's r}}} \\
\cline{2-5} 
& TRI & SNL & TRI & SNL \\
\hline
Double Bacon-Watts model & 0.994 & 0.125 & 0.977 & 0.180 \\
\hline
Proposed method & 1.000 & 0.710 & 0.992 & 0.712 \\
\hline
\end{tabular}
\label{tab3}
\end{adjustbox}
\end{table*}

\begin{figure*}[!ht]
\centering
\subfloat[The time series approximated curvature (top) and corresponding arc curve (bottom).]{\includegraphics[width=2.5in]{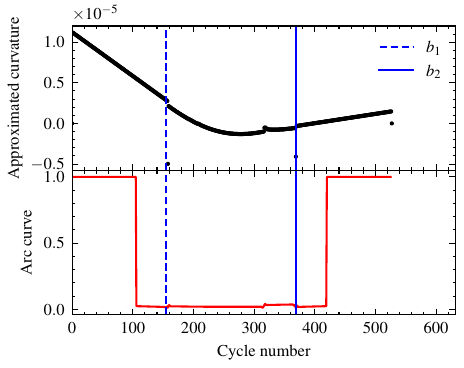}%
\label{fig5_1}}
\hfil
\subfloat[Knee-onset and knee identified with the double Bacon-Watts model and proposed method]{\includegraphics[width=2.5in]{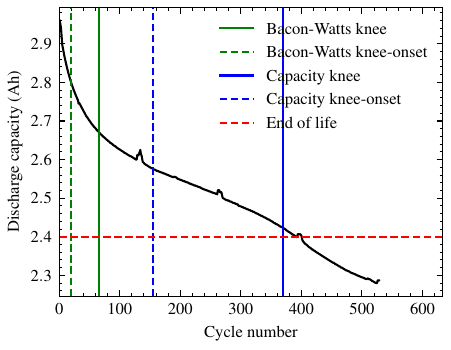}%
\label{fig5_2}}
\caption{The knee-onset and knee identification results of a sample cell [No.10] in the Sandia National Lab dataset.}
\label{fig5}
\end{figure*}

In order to find out the possible cause of the weak correlations between both knee-onsets and knees, identified using the double Bacon-Watts model and end of life, we compared the knee-onsets and the knees obtained using the double Bacon-Watts model and the proposed capacity knee identification method for a sample NMC cell [No.10] in the SNL dataset. 
As illustrated in Fig. \ref{fig5_2}, the sample cell exhibits convex capacity fade in the first degradation phase (i.e., from the beginning of life to the knee-onset point) instead of relatively linear capacity fade in Fig. \ref{fig2_2}. 
The double Bacon-Watts model failed to identify both knee and knee-onset [green lines], while our proposed method successfully identified both knee and knee-onset [blue lines] on this NMC cell. 

The reason for the weak correlations between both knee-onsets and knees, identified using the double Bacon-Watts model and end of life, is that by setting the initial values of the model parameters in Table \ref{tab1}, assumptions regarding the capacity fade curve hold for all the LFP cells in the TRI dataset but not for all the NMC cells in the SNL dataset. In fact, the NMC cells in the SNL dataset exhibit highly diverse capacity fade curves. It is therefore not possible to find one set of initial values in Table \ref{tab1} for all the NMC cells in the SNL dataset. It might be possible to improve the knee-onset and knee identification performance using the double Bacon-Watts model in the SNL dataset by first classifying the NMC cells into several groups based on the similarity level of their capacity fade curves and then setting different initial values in Table \ref{tab1} for each group of NMC cells. However, this will inevitably incur additional work. In contrast, our proposed curvature-based identification method does not require any assumptions of the capacity fade curve, instead, it identifies knee and knee-onset points, given an approximated degradation curvature.

Furthermore, by referring to the diagonal line [solid black line] in Fig. \ref{fig4}, it can be seen that there are NMC cells in the SNL dataset that have knees occurring both before and after the end of life, which motivates the need for classifying retired batteries based on whether or not the knee onset and the knee itself has occurred in their first lives.

\subsubsection{Validation on synthetic dataset}
In order to verify the effectiveness of our proposed method to identify a capacity knee caused by particle cracking, we apply the method to the synthetic battery degradation data generated for the three LGM50 cells with cracking rates as in Table \ref{tab4}. 
It can be seen from Table \ref{tab4} that both knee-onset and knee identified using the proposed method monotonically decrease with increasing cracking rates (from Cell 1 to Cell 3), and we again found a strong linear correlation between knee-onset and knee ($\rho=1.0$) using the proposed knee identification method. Also here, the knee-onset and knee identified using the double Bacon-Watts model show only a weak correlation ($\rho=0.213$).

\begin{table*}[!htbp]
\caption{Battery capacity knee and knee-onset identification performance}
\begin{adjustbox}{width=\textwidth,center}
\begin{tabular}{|l|c|c|c|c|c|c|c|}
\hline
\textbf{Method} & \multicolumn{3}{|c|}{\textbf{Knee (cycles)}} & \multicolumn{3}{|c|}{\textbf{Knee-onset (cycles)}} & \shortstack{\textbf{Identification performance} \\ \textbf{measured by Pearson's r}} \\
\cline{2-7} & \shortstack{Cell 1 \\ (10 $\times$)} & \shortstack{Cell 2 \\ (30 $\times$)} & \shortstack{Cell 3 \\ (50 $\times$)} & \shortstack{Cell 1 \\ (10 $\times$)} & \shortstack{Cell 2 \\ (30 $\times$)} & \shortstack{Cell 3 \\ (50 $\times$)} & \\
\hline
Double Bacon-Watts model & 863 & 804 & 541 & 164 & 537 & 364 & 0.213\\
\hline
Proposed method & 840 & 686 & 455 & 358 & 291 & 191 & 1.000\\
\hline
\multicolumn{8}{l}{\shortstack[l]{Three cells with their cracking rates in Paris' law being 10, 30, and 50 times the standard particle cracking rate \\(i.e., $3.9 \times 10^{-20}$ \cite{purewal2014degradation})}}
\end{tabular}
\label{tab4}
\end{adjustbox}
\end{table*}

\begin{figure*}[!ht]
\centering
\subfloat[The time series approximated curvature (top) and corresponding arc curve (bottom).]{\includegraphics[width=2.5in]{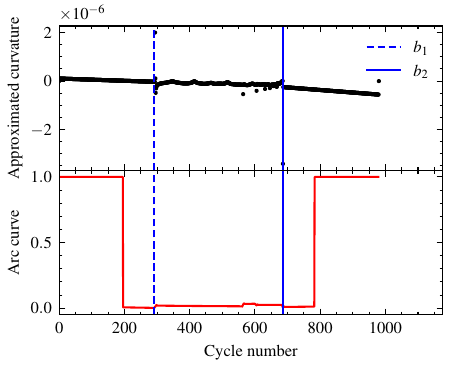}%
\label{fig6_1}}
\hfil
\subfloat[Knee-onset and knee identified with the double Bacon-Watts model and proposed method.]{\includegraphics[width=2.5in]{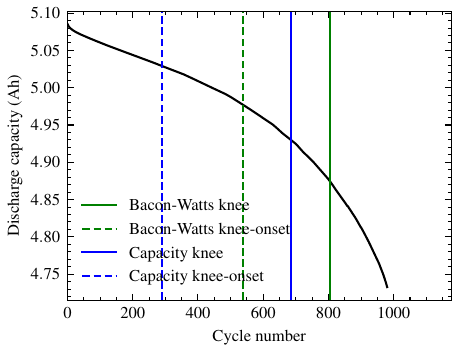}%
\label{fig6_2}}
\caption{The knee-onset and knee identification results of the cylindrical LGM50 cell 2.}
\label{fig6}
\end{figure*}

As a measure of the rate of change of degradation rate, the approximated curvature together with its corresponding CAC is plotted versus the cycle number for the synthetic LGM50 cell 2 in Fig. \ref{fig6_1}. A significant fluctuation of the approximated curvature can again be observed in the second state of which the boundaries [blue lines] were successfully inferred by the proposed identification method. With a state change at each boundary, very few arcs cross over the boundary as most of the approximated curvature subsequences should find their nearest neighbors within the same state. Therefore, the height of the CAC should be the lowest at each boundary where the LGM50 cell degradation process changes from one state to another, as shown at the bottom of Fig. \ref{fig6_1}. 
The measured discharge capacity fade curves are shown in Fig. \ref{fig6_2}. It can be seen that the knee is observable after approximately 680 cycles, and is followed by a sudden failure at 980 cycles due to the fact that the porosity at the negative electrode-separator interface reached zero.

\begin{figure}[!ht]
\centerline{\includegraphics[width=2.5in]{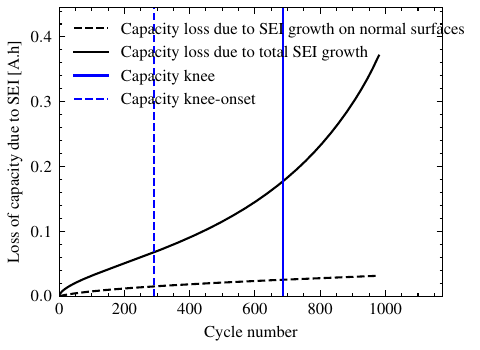}}
\caption{Capacity loss due to SEI growth on the normal particle surface and on the crack surface.}
\label{fig7}
\end{figure}

The loss of capacity is solely caused by LLI due to SEI growth on the normal particle surface and on the cracked surfaces. For a large cracking rate (i.e., 30 times the standard particle cracking rate \cite{purewal2014degradation}), the LLI begins with a square root dependence on time until it reaches the inflection point, which is close to the identified capacity knee-onset point, as shown in Fig. \ref{fig7}. After the inflection point, the LLI accelerates exponentially as the cracks propagate. The accelerated LLI as the internal state change is the only cause for the simulated knee occurrence in this case. The use of synthetic data has the potential for further analysis into the interactions between degradation mechanisms and the evolution of degradation modes behind the observed knee phenomena. Other degradation pathways that may consist of state trajectories of multiple degradation modes can also lead to a knee on the capacity fade curve - knee pathways \cite{attia2022knees}. Therefore, to enable battery degradation diagnosis including knee identification, a larger synthetic dataset that covers a range of other knee pathways needs to be generated, or found in the literature, for example, the Hawaii Natural Energy Institute (HNEI) synthetic dataset by Dubarry and Beck \cite{dubarry2020big} \cite{dubarry2021analysis}.

Overall, the inconsistent results of using the state-of-the-art model (i.e., double Bacon-Watts model) to identify knee-onset and knee on two experimental battery degradation datasets and one synthetic dataset indicate a lack of generalizability of the double Bacon-Watts model towards various battery chemistries for a wide range of operating conditions. In contrast, the generalizability of our proposed capacity knee identification method has been demonstrated on three battery chemistry types, two experimental degradation datasets, and one synthetic dataset, under a wide range of operating conditions. Moreover, as knee identification is formulated as an unsupervised learning problem, knee labeling is not required beforehand. Exact knee-onset and knee points on the capacity fade curve are identified using capacity data as the only input, which is preferable in practical applications, such as a systematic evaluation of the knee prediction performance of both model-based methods and data-driven methods, and facilitating classification of retired electric vehicle batteries from safety and performance perspectives.

\subsection{An application case study: knee-onset early prediction}
As shown in Fig. \ref{fig3} and \ref{fig4}, both knee and knee-onset points identified using our proposed method show strong correlations with the end of life. Additionally, compared to the knee alone, the knee-onset can give a much earlier warning of accelerated degradation (note that on average there are 323 cycles for the LFP cells and 280 cycles for the NMC cells between the knee-onset and the identified knee). Each charge-discharge cycle takes 50 min on average, and consequently, a reduction of 323 cycles in the TRI dataset translates to a reduction of experimental time by approximately 269 hours. Therefore, learning an early-prediction model for battery knee-onset prediction requires much less degradation data than that for battery knee prediction, which in turn significantly reduces experimental time and its associated costs.

\subsubsection{Feature engineering}
To develop machine learning models for battery knee-onset early prediction, input features that are extracted from early degradation data determine their prediction performance. It has been demonstrated by Severson et al. \cite{severson2019data} that a 6-feature set (see Table. \ref{tab5}) results in an early-prediction model with the best battery lifetime prediction performance among the 3 feature sets investigated (i.e., 1-feature set, 6-feature set, and 9-feature set). Therefore, this 6-feature set is used here for training knee-onset early-prediction models and predicting knee-onset points later on. 
\begin{table}[!htbp]
\renewcommand{\arraystretch}{1.5}
\caption{MIT 6-feature set}
\begin{center}
\begin{tabular}{|l|l|}
\hline
\textbf{Feature} & \textbf{Description}\\
\hline
$\mathrm{min}(\Delta Q_{30-10}(V))$ & \shortstack[l]{Minimum of difference of the discharge voltage \\ curve between cycle 30 and cycle 10.}\\
\hline
$\mathrm{var}(\Delta Q_{30-10}(V))$ & \shortstack[l]{Variance of difference of the discharge voltage \\ curve between cycle 30 and cycle 10.}\\
\hline
$\mathrm{ske}(\Delta Q_{30-10}(V))$ & \shortstack[l]{Skewness of difference of the discharge voltage \\ curve between cycle 30 and cycle 10.}\\
\hline
$\mathrm{kur}(\Delta Q_{30-10}(V))$ & \shortstack[l]{Kurtosis of difference of the discharge voltage \\ curve between cycle 30 and cycle 10.}\\
\hline
$Q_2$ & Discharge capacity at cycle 2\\
\hline
$Q_{\mathrm{max}-2}$ & \shortstack[l]{Difference between maximum discharge \\ capacity within the first 30 cycles and \\ discharge capacity at cycle 2}\\
\hline
\end{tabular}
\label{tab5}
\end{center}
\end{table}
\subsubsection{Train-test split}
Most of knee-onset points identified using the proposed method fall between 150 and 270 cycles in the TRI dataset. Thus, these 169 cells are first graded into three classes, i.e., early-knee-onset class ($<150 \ \text{cycles}$), normal-knee-onset class ($150-270 \ \text{cycles}$), and late-knee-onset class ($>270 \ \text{cycles}$). Then, in order to learn a generalized knee-onset early-prediction model, the stratified random sampling method \cite{reitermanova2010data} is employed to randomly split 169 cells, with 80\% in a training set and 20\% in a test set. Equal ratios of early-knee-onset cells, normal-knee-onset cells, and late-knee-onset cells are preserved in the training and test set at each split. Furthermore, the stratified random sampling is repeated 5 times in order to reduce the random effect of the train-test split. The final model performance is averaged over 5 train-test splits.

\subsubsection{Model selection}
By combining a statistical technique called boosting, the Gradient boosting regression trees (GBRT) aggregate a set of "weak" trees to form a single "strong" tree. During the training stage, new trees are generated sequentially to correct the prediction errors of previous trees. This is achieved by minimizing a predefined loss function (e.g., least squares), which quantifies the difference between predicted and actual target values. Moreover, the contribution of each tree to the ensemble model is weighted by the learning rate to prevent overfitting \cite{yang2020lifespan}. 
It has been demonstrated that the GBRT model provided the best battery lifetime early prediction performance among other models (i.e., elastic net, support vector regression, random forests, gaussian process regression, quantile regression forests, quantile regression gradient boosting) using this 6-feature set in previous work \cite{zhang2023comparative}. Since the knee-onset points identified using our proposed method have shown strong correlations with the end of life (see Fig. \ref{fig3}), the GBRT model is selected for knee-onset early prediction in this case study.

\subsubsection{Model performance evaluation}
\begin{figure}[!ht]
\centerline{\includegraphics[width=2.5in]{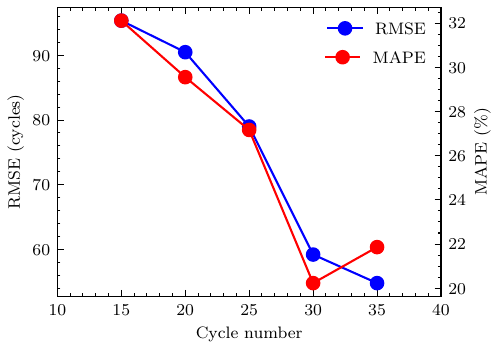}}
\caption{Knee-onset prediction performance as a function of cycle number in the Toyota Research Institute dataset.}
\label{fig8}
\end{figure}

Firstly, a sensitivity analysis of the model performance using different amounts of early degradation data (from the first 15 cycles to the first 35 cycles) is illustrated in Fig. \ref{fig8}. 
It can be seen that the prediction errors, measured by root-mean-square error (RMSE) and mean absolute percentage error (MAPE), significantly decrease with increasing number of cycles. The lowest prediction errors are obtained using degradation data from the first 30 cycles, i.e., RMSE of 59.2 cycles and MAPE of 20.2\%, after which the prediction errors do not decrease significantly. Thus, this model is capable of providing knee-onset prediction with high accuracy after only 30 cycles.

\begin{figure*}[!ht]
\centerline{\includegraphics[width=5in]{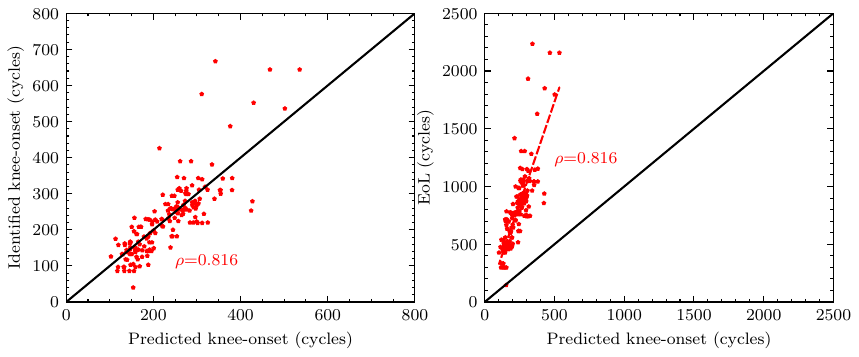}}
\caption{The relationship between predicted knee-onset (using the first 30 cycles data) and identified knee-onset (left), predicted knee-onset and end of life (right) for 169 cells in the Toyota Research Institute dataset.}
\label{fig9}
\end{figure*}

To further investigate the outcomes of early knee-onset prediction using the proposed identification method and its correlation with the battery's end of life, we illustrate the relationship between predicted knee-onset (using the first 30 cycles data) and identified knee-onset, predicted knee-onset and end of life in Fig. \ref{fig9}. Once more there are strong linear correlations between predicted knee-onset and knee-onset identified using our proposed method ($\rho=0.816$), and between predicted knee-onset and end of life ($\rho=0.816$).

Moreover, compared to battery lifetime early prediction using the first 100 cycles data in Severson et al. \cite{severson2019data}, a reduction of 70 cycles in battery knee-onset early prediction translates to a reduction of experimental time by approximately 58 hours, which in turn reduces experimental costs.
\section{Conclusions}
Throughout our literature review, for various chemistries and different operating conditions, we find knees indicating the beginning of accelerated degradation and possible safety issues to occur within a window of 70-95 \% of the initial nominal capacity in experimental testing of commercial lithium-ion cells. To prepare for a successful second-life battery market, concerns arising from possible knee occurrence during first-life and second-life need to be addressed at the repurposing stage.

As the first step to address such concerns, the root causes for the formation of battery capacity knees have been considered and a curvature-based method to identify capacity knee and capacity knee-onset from the capacity fade curve is proposed. By analysis of the approximated curvature in discrete time, a new degradation phenomenon was found. The capacity knee-onset and capacity knee were identified as the start and the end points, respectively, of a transition state of the degradation process, where the approximated curvature fluctuates significantly.
The proposed capacity knee identification method was benchmarked to the state-of-the-art knee identification method (i.e., the double Bacon-Watts model) on both synthetic degradation data and experimental degradation data of both LFP and NMC cells. In the results for the NMC cells, it was demonstrated that the state-of-the-art method failed to identify the knee on the capacity fade curve while our proposed capacity knee identification method successfully identified the knee. The results of the capacity knee identification on synthetic degradation data further validated the effectiveness of our proposed capacity knee identification method.
In contrast to capacity knee identification alone, the knee-onset can give a much earlier warning of accelerated degradation (an average of 323 cycles was found for the LFP cells and an average of 280 cycles for the NMC cells between the knee-onset and the knee). Therefore, learning early-prediction models for battery knee-onset prediction can therefore significantly reduce experimental time and its associated costs. Furthermore, knee-onset prediction can have significant economic value in various industrial applications, such as battery grading, battery replacement planning, and battery repurposing to second-life applications. The capacity knee-onsets and capacity knees that are identified can also be used to systematically evaluate the knee-related prediction performance of both model-based methods and data-driven methods.

Our proposed capacity knee identification method has been validated on battery degradation data from static cycling tests. For wider applications, the method needs to be evaluated on dynamic cycling test data, such as realistic driving profiles for electric vehicles. Then we may not have access to all the cells, but at least the cell with the lowest capacity should be accessible. Therefore it would be recommended to validate our proposed capacity knee identification method on capacity fade data that is real-time estimated in the field in a next step. 
The synthetic dataset of NMC cells with cracking-induced knees has been generated for validation purposes. The use of this synthetic dataset has demonstrated the potential of further analysis into the interactions between degradation mechanisms and the evolution of degradation modes that are associated with knee occurrence. Besides, the knee occurrence can also depend on other degradation pathways that may consist of state trajectories of multiple degradation modes. Hence, to enable online battery degradation diagnosis including knee detection, a larger synthetic dataset that covers a wide range of different knee pathways is needed.
Lastly, considering the close relationship between capacity knees and resistance/impedance elbows, it would also be interesting to adapt the proposed capacity knee identification method accordingly so that the internal resistance elbows can be identified and predicted from internal resistance/impedance data.

\section*{CRediT authorship contribution statement}
\textbf{Huang Zhang:} Conceptualization, Methodology, Software, Validation, Formal analysis, Data curation, Writing – original draft. \textbf{Faisal Altaf:} Resources, Writing – review \& editing, Supervision, Project administration, Funding acquisition. \textbf{Torsten Wik:} Resources, Writing – review \& editing, Supervision, Funding acquisition.

\section*{Declaration of competing interest}
 The authors declare that they have no known competing financial interests or personal relationships that could have appeared to influence the work reported in this paper.

\section*{Acknowledgements}
This work was supported by the Swedish Energy Agency and Volvo Group (Grant number 45540-1). In particular, the authors would like to thank Xixi Liu from the Department of Electrical Engineering, Chalmers University of Technology for her constructive discussions.


\bibliographystyle{elsarticle-num} 
\bibliography{References}

\begin{thebibliography}{10}
\expandafter\ifx\csname url\endcsname\relax
  \def\url#1{\texttt{#1}}\fi
\expandafter\ifx\csname urlprefix\endcsname\relax\def\urlprefix{URL }\fi
\expandafter\ifx\csname href\endcsname\relax
  \def\href#1#2{#2} \def\path#1{#1}\fi

\bibitem{schmuch2018performance}
R.~Schmuch, R.~Wagner, G.~H{\"o}rpel, T.~Placke, M.~Winter, Performance and cost of materials for lithium-based rechargeable automotive batteries, Nature Energy 3~(4) (2018) 267--278.

\bibitem{martinez2018technical}
E.~Martinez-Laserna, E.~Sarasketa-Zabala, I.~V. Sarria, D.-I. Stroe, M.~Swierczynski, A.~Warnecke, J.-M. Timmermans, S.~Goutam, N.~Omar, P.~Rodriguez, Technical viability of battery second life: A study from the ageing perspective, IEEE Transactions on Industry Applications 54~(3) (2018) 2703--2713.

\bibitem{dubarry2012synthesize}
M.~Dubarry, C.~Truchot, B.~Y. Liaw, Synthesize battery degradation modes via a diagnostic and prognostic model, Journal of power sources 219 (2012) 204--216.

\bibitem{he2011prognostics}
W.~He, N.~Williard, M.~Osterman, M.~Pecht, Prognostics of lithium-ion batteries based on dempster--shafer theory and the bayesian monte carlo method, Journal of Power Sources 196~(23) (2011) 10314--10321.

\bibitem{yang2017prognostics}
F.~Yang, D.~Wang, Y.~Xing, K.-L. Tsui, Prognostics of li (nimnco) o2-based lithium-ion batteries using a novel battery degradation model, Microelectronics Reliability 70 (2017) 70--78.

\bibitem{attia2022knees}
P.~M. Attia, A.~Bills, F.~B. Planella, P.~Dechent, G.~Dos~Reis, M.~Dubarry, P.~Gasper, R.~Gilchrist, S.~Greenbank, D.~Howey, et~al., “knees” in lithium-ion battery aging trajectories, Journal of The Electrochemical Society 169~(6) (2022) 060517.

\bibitem{martinez2016evaluation}
E.~Martinez-Laserna, E.~Sarasketa-Zabala, D.-I. Stroe, M.~Swierczynski, A.~Warnecke, J.-M. Timmermans, S.~Goutam, P.~Rodriguez, Evaluation of lithium-ion battery second life performance and degradation, in: 2016 IEEE Energy Conversion Congress and Exposition (ECCE), IEEE, 2016, pp. 1--7.

\bibitem{zhang2019accelerated}
C.~Zhang, Y.~Wang, Y.~Gao, F.~Wang, B.~Mu, W.~Zhang, Accelerated fading recognition for lithium-ion batteries with nickel-cobalt-manganese cathode using quantile regression method, Applied Energy 256 (2019) 113841.

\bibitem{wood2011investigation}
E.~Wood, M.~Alexander, T.~H. Bradley, Investigation of battery end-of-life conditions for plug-in hybrid electric vehicles, Journal of Power Sources 196~(11) (2011) 5147--5154.

\bibitem{arrinda2021application}
M.~Arrinda, M.~Oyarbide, H.~Macicior, E.~Muxika, H.~Popp, M.~Jahn, B.~Ganev, I.~Cendoya, Application dependent end-of-life threshold definition methodology for batteries in electric vehicles, Batteries 7~(1) (2021) 12.

\bibitem{baumann2018parameter}
M.~Baumann, L.~Wildfeuer, S.~Rohr, M.~Lienkamp, Parameter variations within li-ion battery packs--theoretical investigations and experimental quantification, Journal of Energy Storage 18 (2018) 295--307.

\bibitem{ahmadi2017cascaded}
L.~Ahmadi, S.~B. Young, M.~Fowler, R.~A. Fraser, M.~A. Achachlouei, A cascaded life cycle: reuse of electric vehicle lithium-ion battery packs in energy storage systems, The International Journal of Life Cycle Assessment 22~(1) (2017) 111--124.

\bibitem{diao2019algorithm}
W.~Diao, S.~Saxena, B.~Han, M.~Pecht, Algorithm to determine the knee point on capacity fade curves of lithium-ion cells, Energies 12~(15) (2019) 2910.

\bibitem{fermin2020identification}
P.~Ferm{\'\i}n-Cueto, E.~McTurk, M.~Allerhand, E.~Medina-Lopez, M.~F. Anjos, J.~Sylvester, G.~Dos~Reis, Identification and machine learning prediction of knee-point and knee-onset in capacity degradation curves of lithium-ion cells, Energy and AI 1 (2020) 100006.

\bibitem{greenbank2021automated}
S.~Greenbank, D.~Howey, Automated feature extraction and selection for data-driven models of rapid battery capacity fade and end of life, IEEE Transactions on Industrial Informatics 18~(5) (2021) 2965--2973.

\bibitem{sohn2022two}
S.~Sohn, H.-E. Byun, J.~H. Lee, Two-stage deep learning for online prediction of knee-point in li-ion battery capacity degradation, Applied Energy 328 (2022) 120204.

\bibitem{costa2024icformer}
N.~Costa, D.~Anse{\'a}n, M.~Dubarry, L.~S{\'a}nchez, Icformer: A deep learning model for informed lithium-ion battery diagnosis and early knee detection, Journal of Power Sources 592 (2024) 233910.

\bibitem{yeh2016matrix}
C.-C.~M. Yeh, Y.~Zhu, L.~Ulanova, N.~Begum, Y.~Ding, H.~A. Dau, D.~F. Silva, A.~Mueen, E.~Keogh, Matrix profile i: all pairs similarity joins for time series: a unifying view that includes motifs, discords and shapelets, in: 2016 IEEE 16th international conference on data mining (ICDM), Ieee, 2016, pp. 1317--1322.

\bibitem{gharghabi2017matrix}
S.~Gharghabi, Y.~Ding, C.-C.~M. Yeh, K.~Kamgar, L.~Ulanova, E.~Keogh, Matrix profile viii: domain agnostic online semantic segmentation at superhuman performance levels, in: 2017 IEEE international conference on data mining (ICDM), IEEE, 2017, pp. 117--126.

\bibitem{satopaa2011finding}
V.~Satopaa, J.~Albrecht, D.~Irwin, B.~Raghavan, Finding a" kneedle" in a haystack: Detecting knee points in system behavior, in: 2011 31st international conference on distributed computing systems workshops, IEEE, 2011, pp. 166--171.

\bibitem{raj2020investigation}
T.~Raj, A.~A. Wang, C.~W. Monroe, D.~A. Howey, Investigation of path-dependent degradation in lithium-ion batteries, Batteries \& Supercaps 3~(12) (2020) 1377--1385.

\bibitem{bach2016nonlinear}
T.~C. Bach, S.~F. Schuster, E.~Fleder, J.~M{\"u}ller, M.~J. Brand, H.~Lorrmann, A.~Jossen, G.~Sextl, Nonlinear aging of cylindrical lithium-ion cells linked to heterogeneous compression, Journal of Energy Storage 5 (2016) 212--223.

\bibitem{severson2019data}
K.~A. Severson, P.~M. Attia, N.~Jin, N.~Perkins, B.~Jiang, Z.~Yang, M.~H. Chen, M.~Aykol, P.~K. Herring, D.~Fraggedakis, et~al., Data-driven prediction of battery cycle life before capacity degradation, Nature Energy 4~(5) (2019) 383--391.

\bibitem{savitzky1964smoothing}
A.~Savitzky, M.~J. Golay, Smoothing and differentiation of data by simplified least squares procedures., Analytical chemistry 36~(8) (1964) 1627--1639.

\bibitem{scheffer2009early}
M.~Scheffer, J.~Bascompte, W.~A. Brock, V.~Brovkin, S.~R. Carpenter, V.~Dakos, H.~Held, E.~H. Van~Nes, M.~Rietkerk, G.~Sugihara, Early-warning signals for critical transitions, Nature 461~(7260) (2009) 53--59.

\bibitem{liu2015identifying}
R.~Liu, P.~Chen, K.~Aihara, L.~Chen, Identifying early-warning signals of critical transitions with strong noise by dynamical network markers, Scientific reports 5~(1) (2015) 17501.

\bibitem{attia2020closed}
P.~M. Attia, A.~Grover, N.~Jin, K.~A. Severson, T.~M. Markov, Y.-H. Liao, M.~H. Chen, B.~Cheong, N.~Perkins, Z.~Yang, et~al., Closed-loop optimization of fast-charging protocols for batteries with machine learning, Nature 578~(7795) (2020) 397--402.

\bibitem{lewerenz2017post}
M.~Lewerenz, A.~Warnecke, D.~U. Sauer, Post-mortem analysis on lifepo4| graphite cells describing the evolution \& composition of covering layer on anode and their impact on cell performance, Journal of Power Sources 369 (2017) 122--132.

\bibitem{ansean2017operando}
D.~Anse{\'a}n, M.~Dubarry, A.~Devie, B.~Liaw, V.~Garc{\'\i}a, J.~Viera, M.~Gonz{\'a}lez, Operando lithium plating quantification and early detection of a commercial lifepo4 cell cycled under dynamic driving schedule, Journal of Power Sources 356 (2017) 36--46.

\bibitem{lewerenz2017differential}
M.~Lewerenz, A.~Marongiu, A.~Warnecke, D.~U. Sauer, Differential voltage analysis as a tool for analyzing inhomogeneous aging: A case study for lifepo4| graphite cylindrical cells, Journal of Power Sources 368 (2017) 57--67.

\bibitem{preger2020degradation}
Y.~Preger, H.~M. Barkholtz, A.~Fresquez, D.~L. Campbell, B.~W. Juba, J.~Rom{\`a}n-Kustas, S.~R. Ferreira, B.~Chalamala, Degradation of commercial lithium-ion cells as a function of chemistry and cycling conditions, Journal of The Electrochemical Society 167~(12) (2020) 120532.

\bibitem{ma2019hindering}
X.~Ma, J.~E. Harlow, J.~Li, L.~Ma, D.~S. Hall, S.~Buteau, M.~Genovese, M.~Cormier, J.~Dahn, Hindering rollover failure of li [ni0. 5mn0. 3co0. 2] o2/graphite pouch cells during long-term cycling, Journal of The Electrochemical Society 166~(4) (2019) A711--A724.

\bibitem{pfrang2018long}
A.~Pfrang, A.~Kersys, A.~Kriston, D.~Sauer, C.~Rahe, S.~K{\"a}bitz, E.~Figgemeier, Long-term cycling induced jelly roll deformation in commercial 18650 cells, Journal of Power Sources 392 (2018) 168--175.

\bibitem{willenberg2020high}
L.~K. Willenberg, P.~Dechent, G.~Fuchs, D.~U. Sauer, E.~Figgemeier, High-precision monitoring of volume change of commercial lithium-ion batteries by using strain gauges, Sustainability 12~(2) (2020) 557.

\bibitem{doyle1995use}
M.~Doyle, J.~Newman, The use of mathematical modeling in the design of lithium/polymer battery systems, Electrochimica Acta 40~(13-14) (1995) 2191--2196.

\bibitem{sulzer2021python}
V.~Sulzer, S.~G. Marquis, R.~Timms, M.~Robinson, S.~J. Chapman, Python battery mathematical modelling (pybamm), Journal of Open Research Software 9~(1) (2021).

\bibitem{chen2020development}
C.-H. Chen, F.~B. Planella, K.~O’regan, D.~Gastol, W.~D. Widanage, E.~Kendrick, Development of experimental techniques for parameterization of multi-scale lithium-ion battery models, Journal of The Electrochemical Society 167~(8) (2020) 080534.

\bibitem{o2022lithium}
S.~E. O'Kane, W.~Ai, G.~Madabattula, D.~Alonso-Alvarez, R.~Timms, V.~Sulzer, J.~S. Edge, B.~Wu, G.~J. Offer, M.~Marinescu, Lithium-ion battery degradation: how to model it, Physical Chemistry Chemical Physics 24~(13) (2022) 7909--7922.

\bibitem{purewal2014degradation}
J.~Purewal, J.~Wang, J.~Graetz, S.~Soukiazian, H.~Tataria, M.~W. Verbrugge, Degradation of lithium ion batteries employing graphite negatives and nickel--cobalt--manganese oxide+ spinel manganese oxide positives: Part 2, chemical--mechanical degradation model, Journal of power sources 272 (2014) 1154--1161.

\bibitem{dubarry2020big}
M.~Dubarry, D.~Beck, Big data training data for artificial intelligence-based li-ion diagnosis and prognosis, Journal of Power Sources 479 (2020) 228806.

\bibitem{dubarry2021analysis}
M.~Dubarry, D.~Beck, Analysis of synthetic voltage vs. capacity datasets for big data li-ion diagnosis and prognosis, Energies 14~(9) (2021) 2371.

\bibitem{reitermanova2010data}
Z.~Reitermanova, et~al., Data splitting, in: WDS, Vol.~10, 2010, pp. 31--36.

\bibitem{yang2020lifespan}
F.~Yang, D.~Wang, F.~Xu, Z.~Huang, K.-L. Tsui, Lifespan prediction of lithium-ion batteries based on various extracted features and gradient boosting regression tree model, Journal of Power Sources 476 (2020) 228654.

\bibitem{zhang2023comparative}
H.~Zhang, F.~Altaf, T.~Wik, S.~Gros, Comparative analysis of battery cycle life early prediction using machine learning pipeline, IFAC-PapersOnLine 56~(2) (2023) 3757--3763.

\end{thebibliography}





\end{document}